\documentclass[aps,pra,preprint,showpacs,amsmath,amssymb,groupedaddress]{revtex4}
\usepackage{bm}
\def\sg{\sigma}
\def\al{\alpha}
\def\s12{\sqrt{a_1a_2}}
\def\f{\sqrt{\frac{a_1}{a_{12}}}}
\def\g{\sqrt{\frac{a_2}{a_{12}}}}
\def\aa#1#2#3#4{{((#1\al #2)\al #3)}_{#4}}
\def\ssl#1#2#3#4{{((#1\sg #2)\sg #3)}_{#4}}
\def\as#1#2#3#4{{((#1\al #2)\sg #3)}_{#4}}
\def\sa#1#2#3#4{{((#1\sg #2)
\al #3)}_{#4}}

\def\ssr#1#2#3#4{{(#1\sg (#2\sg #3))}_{#4}}

\def\sar#1#2#3#4{{(#1\sg (#2\al #3))}_{#4}}
\begin{document}

\title{Mixing quantum and classical mechanics and uniqueness of
  Planck's constant} \author{Debendranath Sahoo}
\affiliation{Institute of Physics, P.O. Sainik School, Bhubaneswar 751
  005, Orissa, India} \email[]{dsahoo@iopb.res.in}
\altaffiliation{Permanent address: Materials Science Division, Indira
  Gandhi Centre for Atomic Research, Kalpakkam, Tamil Nadu, PIN: 603
  102, India.}
\begin{abstract}
  Observables of quantum or classical mechanics form algebras called
  quantum or classical Hamilton algebras respectively
    (Grgin E and Petersen A (1974) {\it J Math
    Phys} {\bf 15} 764\cite{grginpetersen}, Sahoo D (1977) {\it
    Pramana} {\bf 8} 545\cite{sahoo}).  We show that the
  tensor-product of two quantum Hamilton algebras, each characterized
  by a different Planck's constant is an algebra of the same type
  characterized by yet another Planck's constant. The algebraic
  structure of mixed quantum and classical systems is then analyzed by
  taking the limit of vanishing Planck's constant in one of the
  component algebras. This approach provides new insight into failures
  of various formalisms dealing with mixed quantum-classical systems.
  It shows that in the interacting mixed quantum-classical
  description, there can be no back-reaction of the quantum system on
  the classical. A natural algebraic requirement involving restriction
  of the tensor product of two quantum Hamilton algebras to their
  components proves that Planck's constant is unique.
\end{abstract}

\maketitle

\section{Introduction}
\label{sec:Intro}
The title of the present paper may create an impression that two
disjoint subjects are being discussed together. However, a little
reflection would convince the reader that there is a connecting thread
between the two. Inherent in a quantum system is a Planck's constant
(PC) governing its behaviour whereas a classical system can be thought
of as a system with zero PC.  Thus in mixed quantum-classical
mechanics, we are dealing with two systems with different PC's. Now in
proving the uniqueness of PC, it is but natural to consider two
systems with different values of PC and then to examine the
consequences. The fact that two different PC's come into play in the
analysis of both the subjects, provides the connecting thread. The
purpose of this work is two-fold: to investigate {\em why one can not
  have a fundamentally satisfactory dynamical description of
  interacting quantum-classical systems} and {\em to understand the
  uniqueness of PC} conventionally assumed in the physics literature.
The method of our investigation is of algebraic nature.

As of now there is no consistent theory of interaction of a classical
system with a quantum one. Such a theory is desirable since a variety
of problems in a number of different fields involve coupling of
quantum and classical degrees of freedom. In the development of
quantum mechanics (QM), Niels Bohr\cite{bohr} had always insisted that
measuring instruments must be describable in classical terms, but did
not provide a theoretical framework for the description of interacting
quantum-classical systems. The so-called Copenhagen interpretation of
quantum theory, founded on this assumption, is by and large accepted
by all physicists; yet this lacuna has remained as a sore point. The
issue of a mixed quantum-classical description is important in the
discussion of early universe physics where fully quantum matter fields
have to be necessarily coupled to the gravitational field which is
classical. Traditional approach to this problem has been to couple the
gravity field to the expectation values of the quantum energy-momentum
tensor of the matter fields. In this approach one misses the effects
of quantum fluctuations on the classical gravitational field--the
so-called quantum back-reaction.

There has been no dearth of effort in constructing a mathematically
consistent theory of such mixed systems.  Some
  authors\cite{anderson,bouchertraschen,salcedo,carosalcedo} use mixed
  classical-quantum notation to denote the dynamical variables (DV's)
  of the (mixed) system. Let $x,y,x_i,\cdots$ denote classical DV's
  which are ordinary functions of commuting phase-space variables and
  let $X,Y,X_i,\cdots$ denote quantum DV's which are noncommuting
  operators (acting on some suitable Hilbert space of states.
  Then typical DV's of the mixed
  quantum-classical system are denoted as $Xx,Yy,\cdots$, which are
  operator-valued functions. Also the general DV's of the mixed system
  may be denoted ${\cal X,Y,}\cdots$ where ${\cal X}=\sum_i X_ix_i$,
  ${\cal Y}=\sum_i Y_iy_i$, etc. Let us use the notation
\begin{eqnarray}
{[X,Y]}^- &=& \frac{(XY-YX)}{i\hbar},\\
{[X,Y]}^+ &=& \frac{1}{2}(XY+YX),
\end{eqnarray}
to denote the commutator and the anticommutator brackets respectively
and let ${\{x,y\}}_P$ denote the usual Poisson bracket of classical
DV's. In order to denote the corresponding bracket of mixed DV's, let
us adopt the notation $[\{ , \}]$. Guided mainly by guess-work,
Following definitions of this bracket have been proposed:
\begin{eqnarray}
\label{BT}
&&[\{xX,yY\}]=xy{[X,Y]}^-+{\{x,y\}}_P{[X,Y]}^+\nonumber\\
&&\mbox{(Boucher and
  Traschen\cite{bouchertraschen})}\\
\label{BT1}
&&[\{{\cal X},{\cal Y}\}]=[{\cal X},{\cal Y}]+
\frac{1}{2}\left( {\{ {\cal X},{\cal Y}\}}_P-{\{ {\cal Y},{\cal X}\}}_P\right)\nonumber\\
&&\mbox{(Aleksandrov\cite{aleksandrov}, Caro and Salcedo\cite{carosalcedo})}\\
&&[\{{\cal X},{\cal Y}\}]=[{\cal X},{\cal Y}]+{\{{\cal X},{\cal Y}\}}_P\quad
\mbox{(Anderson\cite{anderson})}
\label{ander}
\end{eqnarray}
 Caro and Salcedo\cite{carosalcedo} consider a quantum system
consisting of two mutually interacting subsystems and enquire whether
it is possible to take the classical limit ({\em i\ e\ ,} letting $\hbar \to
0$) in just one of the subsystems maintaining at the same time an
internally consistent dynamics for the resulting mixed
quantum-classical system. They call this as the {\it semi-quantization
  problem} and arrive at the bracket (\ref{BT1}). Further, they show
that this bracket, although antisymmetric, does {\em not} satisfy the
Jacobi identity. In his investigation of quantum back-reaction on
classical variables, Anderson\cite{anderson} suggests the bracket
(\ref{ander}), which is not even antisymmetric.  Prezhdo and
Kisil\cite{prezhdokisil} develop a mathematically sophisticated
formalism and arrive at the result identical to (\ref{BT}) but written
in terms of the symbols of the operators (see their Eq.(24)).

A satisfactory bracket ( , ) describing any dynamics, be it classical
or quantum (and also desirable for a mixed classical-quantum system),
must possess the following properties\cite{dirac}:
\begin{eqnarray}
&&(A,B)=-(B,A) \quad\mbox{(antisymmetry)},\\
&&((A,B),C)+((B,C),A)+((C,A),B)=0\quad\mbox{(Jacobi identity)},\\
&&(A,BC)=(A,B)C+B(A,C)\quad\mbox{(Derivation identity)}.
\end{eqnarray}
As is well known\cite{carosalcedo}, antisymmetry of the bracket
ensures conservation of energy, the Jacobi identity ensures that
$(A,B)$ also evolves dynamically and the derivation identity ensures
that the product $AB$ also evolves with consistent dynamics. Lack of
any of these properties impose severe impediments to mixing of
classical and quantum degrees of freedom.  The bracket (\ref{ander})
has none of these properties whereas the bracket (\ref{BT}) satisfies
neither the Jacobi nor the derivation
identity\cite{carosalcedo,anderson1}.  Diosi and his group\cite{diosi}
extensively investigate the question of coupling of quantum-classical
systems focussing their attention mainly on maintaining the positivity
of the quantum states.  Hay and Peres\cite{hayperes} treat the
  apparatus quantum mechanically while it interacts with the system
  and then give a classical description of the apparatus within the
  framework of the Wigner functions.
  Peres and Temo\cite{perestemo} develop a hybrid formalism by taking
  recourse to the Koopman operator representation of classical
  Hamiltonians and conclude that the correspondence principle is
  violated due to the interaction.  Belavkin and his
  collaborators\cite{belavkin} develop a stochastic Hamiltonian theory
  for coupling of a quantum system with an apparatus. 
  In this approach, attention is focussed on providing
  purely dynamical arguments to derive entanglement, decoherence and
  collapse of the coupled system consisting of a quantum system and a
  (semi-classical) apparatus.  However, in this theory, as in the
  original von Neumann theory\cite{vonneumann} of measurement, both
  the system and the apparatus are treated quantum mechanically and a
  ``reduction model'' is proposed achieving some improvement over the
  von Neumann reduction postulate. Some more discussion of this model
  will be given in the last section dealing with discussion of our
  results. Sudarshan and his collaborators\cite{sudarshan}
  propose a novel procedure of coupling a classical system (the
  apparatus) with a quantum one. They embed the classical system into
  what they term as a classical enlarged quantum system (CEQS). The
  set of observable phase-space variables (regarded as commuting
  operators) are supplemented by an equal number of unobservable
  conjugate variables (noncommuting with the previous set). In this
  approach, after coupling the CEQS with the quantum system and
  subsequently decoupling these systems, the value of the measured
  quantity of the quantum system is transfered to an appropriate
  observable of the CEQS. Certain principle of integrity invoked by
  these authors, assures satisfactory behaviour of observables of the
  classical system in the measurement process. However, the state
  evolution of the CEQS indicates that the classical system does {\em
    not} remain purely classical after the interaction is over.  The
basic mathematical problem of a truly satisfactory quantum-classical
coupling remains unsolved.  The question whether there exists a
satisfactory bracket for the mixed quantum-classical systems calls for
a detailed investigation from a purely algebraic point of view. The
present work has this as one of its main motivation.

The second important result which is derived in this work is an
algebraic proof that PC is unique. The possibility of multiplicity of
PC's cannot be logically ruled out and its universality is of
empirically established nature\cite{wichman}. Fischback {\it et al}
\cite{fischback} have examined this question carefully and after
recalling how the existence of several PC's leads to violation of
space-time symmetry laws, suggest a possible test for experimentally
verifying this assumption. Battaglia\cite{battaglia}, while arguing
that the introduction of several PC's is undesirable, suggest remedies
to some of the embarrassing problems arising in the event that
experiments do allow for such an eventuality.  Our proof of uniqueness
of PC resolves this issue. We now describe the abstract algebraic
structures of quantum and classical mechanics.

\section{Hamilton algebra}
\noindent
In order to motivate the definition of the algebra of observables of
QM we note that in the von Neumann formulation\cite{vonneumann} of QM,
the associative algebra $\cal B(S)$ of bounded linear operators,
defined over the complex field $\mathbb{C}$, acts on a Hilbert space
$\cal S$ of states. The set of observables $\bar{\cal B}(\cal S)$,
consisting of self-adjoint elements of $\cal B(S)$ and defined over
the real field $\mathbb{R}$ inherits from $\cal B(S)$ the structure of
a Jordan-Lie algebra\cite{emch} with a Jordan product ${[X,Y]}^+$ and
a Lie product ${[X,Y]}^-$ where $X$, $Y\,\in\bar{\cal B}(\cal S)$.
Since every $X\in\cal B(S)$ can be written uniquely in the form
$X=X_1+iX_2$ with $X_1$, $X_2\in\bar{\cal B}(S)$ and $i=\sqrt{-1}$,
$\cal B(S)$ is the complex extension of $\bar{\cal B}(\cal S)$.  This
observation suggests the definition of a quantum Hamilton algebra
(QHA) \cite{sahoo,sahoo1}: it is a two-product algebra $\{{\cal
  H},\alpha ^a ,\sigma ^a ,\mathbb{R}\}$, over the real field
$\mathbb{R}$, parametrized by a real number $a$ called the {\em
  quantum constant}. Here $\cal H$ is the linear space underlying the
algebra; $\alpha ^a$ and $\sigma ^a$ are bilinear products $\alpha
^a$, $\sigma ^a$: ${\cal H}\otimes{\cal H}\to{\cal H}$. Henceforth we
shall denote this algebra by the notation ${\cal H}^a$. The
correspondence between the symbols and their abstract counterparts is:
$\bar{\cal B}(\cal S)\to\cal H$, ${[ , ]}^-\to\alpha ^a$, ${[ ,
  ]}^+\to\sigma ^a$ and $a\to\hbar ^2/4$.  Elements of the set $\cal
H$ are denoted by $e,f,g,h,\cdots$, where $e$ is the unit element of
${\cal H}^a$ (with respect to $\sigma ^a$). Note that in the
algorithmic form, we have earlier used the notation $X$, $Y$, etc. to
denote the (operator) elements of the elements of $\bar{\cal B}(\cal
S)$. A QHA is defined\cite{sahoo,sahoo1} by the identities:
\begin{eqnarray}
&& f\alpha ^a g = - g\alpha ^a f\quad\mbox{(antisymmetry)},
\label{eq:antisym}\\
&& f\alpha ^a (g\alpha ^ah)+g\alpha ^a (h\alpha ^a f)+h\alpha
^a(f\alpha ^a g)=0\quad\mbox{(Jacobi identity)},
\label{eq:jacoby}\\
&& f\sigma ^a g=g\sigma ^a f \quad\mbox{(symmetry)},
\label{eq:sym}\\
&& f\alpha ^a (g\sigma ^a h)=(f\alpha ^a g)\sigma ^a h+g\sigma ^a
(f\alpha ^a h)
\quad\mbox{(derivation of $\alpha ^a$ wrt $\sigma ^a$ )},
\label{eq:der}\\
&& \Delta _\sigma ^a(f,g,h)\equiv (f\sigma ^a g)\sigma ^a h-f\sigma ^a
(g\sigma ^ah)\nonumber\\
&&=a[(f\alpha ^a h)\alpha ^a g]\quad\mbox{(Canonical relation, CR)}.
\label{eq:asso}
\end{eqnarray}
The CR can be trivially checked to hold in its algorithmic form in
$\bar{\cal B}(\cal S)$. Elevation of this trivial looking relation in
this form to the status of a defining identity of our algebra follows
from the composition properties of the HA's which hold if and only if
the CR is assumed to hold.  Note that both $\sigma ^a$ and $\alpha ^a$
are nonassociative products; the nonassociativity of $\sigma ^a$ is
measured by the associator $\Delta _{\sigma ^a}$, a trilinear object;
and the rhs of Eq.(\ref{eq:asso}) can also be written (but for the
constant factor $a$) as an $\alpha ^a$-associator. Thus the CR is an
exact relation between the two associators. Note also that the
standard Jordan identity\cite{schafer} $f^2\sigma ^a (g\sigma ^a
f)=(f^2\sigma ^ag)\sigma ^af$ with $f^2=f\sigma ^a f$ follows from
(\ref{eq:asso}) by substituting in it $f=h$ and using
(\ref{eq:antisym}). It is the interaction of the Lie and the Jordan
structures via Eqs. (\ref{eq:der}) and (\ref{eq:asso}) that makes a
QHA an interesting algebraic object in its own right.

A classical Hamilton algebra (CHA) ${\cal H}^0=\{ {\cal H}, \alpha ^0,
\sigma ^0,\mathbb{R}\}$ is now defined by setting $a=0$ in the
identities (\ref{eq:antisym}-\ref{eq:asso}). Note that the product
$\sigma ^0$ is associative in addition to being commutative--a
property of classical phase space functions.  We next turn our
attention on the most important characteristic of Hamilton algebras
(HA's).

\section{Composition properties of Hamilton algebras}
\noindent
Interaction of two quantum systems should result in a composite system
describable within the same framework. This intuitive idea is made
rigorous by postulating that the tensor-product (TP) composition of
two QHA's is yet another QHA. It can be easily verified that the
auxiliary product
\begin{equation}
\label{eq:tau}
\tau ^a= \sigma ^a+ \sqrt{-a}\,\alpha ^a
\end{equation}
defined in the complex extension ${\cal A}^a$ of ${\cal H}^a$, $\tau
^a$ : ${\cal A}^a\otimes{\cal A}^a\to{\cal A}^a$ is an associative
product. Here the detailed form of the above symbolic relation is
$$
f\tau ^ag = f\sigma ^ag + \sqrt{-a}\, f\alpha ^a
g,\quad\mbox{for}\quad f,g\in{\cal A}^a.
$$
Note that $\sigma ^a$ and $\alpha ^a$ are derived products:
\begin{eqnarray}
f\sigma ^ag=\frac{1}{2}(f\tau ^ag+g\tau ^af),\\
f\alpha ^ag=\frac{1}{2\sqrt{-a}}(f\tau ^ag-g\tau ^af).
\end{eqnarray}
The algebra $\{{{\cal A}, \tau ^a, \mathbb{C}\}}^a={\cal A}^a$, where
$\mathbb{C}$ is the complex field, is the associative envelope algebra
of ${\cal H}^a$. We now follow the standard procedure\cite{greub} of
forming the tensor product ${\cal A}^{a_{12}}$ of ${\cal A}^{a_1}$ and
${\cal A}^{a_2}$. The precise definition of $\tau _{12}$ makes use of
the ``switching map''
\begin{eqnarray}
&&S:({\cal A}_1\otimes{\cal A}_2)\otimes ({\cal A}_1\otimes{\cal A}_2)
\longrightarrow({\cal A}_1\otimes{\cal A}_1)
\otimes({\cal A}_2\otimes{\cal A}_2),\\
&&S:(f_1\otimes f_2)\otimes (g_1\otimes g_2)\mapsto (f_1\otimes g_1)\otimes (f_2\otimes g_2).
\end{eqnarray}
Then one has
\begin{equation}
\tau _{12}=(\tau _1\otimes\tau _2)\circ S,
\label{eq:taucomp}
\end{equation}
Here for brevity we have suppressed the superscripts in the $\tau$'s.
Thus $\tau _1$ stands for $\tau ^{a_1}_1$, etc.  Similar notation will
be used for the other products too. The symbol $\circ$ denotes the
composition of maps.  For brevity we shall use the notation
$f_{12}\equiv f_1\otimes f_2$, ${(f\sigma g)}_{12}\equiv
f_{12}\,{\sigma}_{12}\,g_{12}$,\ ${\left[\Delta _\sigma
    (f,g,h)\right]}_{12}\equiv {(f\sigma g)}_{12}\,{\sigma} _{12}\,
h_{12}-f_{12}\,\sigma _{12} \,{(g\sigma h)}_{12}$, etc.  Consider two
QHA's ${{\cal H}_1}^{a_1}$ and ${{\cal H}_2}^{a_2}$.  Let
\begin{eqnarray}
&&\tau _{12}=\sigma _{12}+\sqrt{-a_{12}}\,\alpha _{12},\\
&&\tau _k=\sigma _k+\sqrt{-a_k}\,\alpha _k\quad\mbox{(k=1,2)}.
\end{eqnarray}
Then Eq.(\ref{eq:taucomp}) implies
\begin{eqnarray}
&&\sigma _{12}+\sqrt{-a_{12}}\,\alpha _{12}
=(\sigma _1+\sqrt{-a_1}\,\alpha _1)\otimes (\sigma
_2+\sqrt{-a_2}\,\alpha _2)\circ S\nonumber\\
&&=\left[(\sg _1\otimes\sg _2-\sqrt{a_1a_2}\,\al _1\otimes\al _2)
+(\sqrt{-a_1}\,\al _1\otimes\sg
_2+\sqrt{-a_2}\,\sg _1\otimes\al _2\right].
\end{eqnarray}
Now equating the symmetric and antisymmetric parts of both sides we
obtain the composition properties of the derived products:
\begin{eqnarray}
&&\sigma _{12}=[(\sigma _1\otimes\sigma _2)-\sqrt{a_1a_2}\,\, (\alpha
_1\otimes\alpha _2)]\circ S,
\label{eq:sig12}\\
&&\alpha _{12}=\left[ \left(\sqrt{\frac{a_1}{a_{12}}}(\alpha _1\otimes\sigma _2)
+\sqrt{\frac{a_2}{a_{12}}}(\sigma _1\otimes\alpha _2)\right) \right] \circ S.
\label{eq:al12}
\end{eqnarray}
Expanded forms of Eqs. (\ref{eq:sig12}) and (\ref{eq:al12}) are
\begin{equation}
\label{newcompsg}
{(f\sigma g)}_{12} = {(f\sigma g)}_1\otimes {(f\sigma g)}_2-\sqrt{a_1a_2}\,\,{(f\alpha
  g)}_1\otimes {(f\alpha g)}_2,
\end{equation}
\begin{equation}
\label{newcompal}
{(f\alpha g)}_{12} = \left[ \left(\sqrt{\frac{a_1}{a_{12}}} {(f\alpha g)}_1\otimes
    {(f\sigma g)}_2+\sqrt{\frac{a_2}{a_{12}}}{(f\sigma g)}_1\otimes
    {(f\alpha g)}_2\right) \right].
\end{equation}
With these composition laws of the algebraic products, we now proceed
to prove that the identities (\ref{eq:antisym}-\ref{eq:asso}) also
hold in the algebra ${\cal H}^{a_{12}}$. This is a {\em highly
  nontrivial} result since the TP of two algebras in general leads to
an algebra not necessarily of the same type. For example, the TP of
two Lie algebras has a product which is symmetric contrary to the
antisymmetric nature of the Lie product.  Let $f_{12}=f_1\otimes f_2$,
$g_{12}=g_1\otimes g_2$ and $h_{12}=h_1\otimes h_2$ denote three
arbitrary elements in ${\cal H}_1\otimes {\cal H}_2$. For brevity we
shall write $f_1\otimes f_2=f_1f_2$, etc.  We now demonstrate the
following results.  \newtheorem{theoremdemo}{Lemma}
\begin{theoremdemo}
  $\alpha _{12}$ is antisymmetric.
\end{theoremdemo}
{\em Proof}:
\begin{eqnarray}
{(f\alpha g)}_{12} &=&
\f {(f\alpha g)}_1{(f\sigma g)}_2+
\g {(f\sigma g)}_1{(f\alpha g)}_2
\nonumber\\
&=&-\f {(g\alpha f)}_1{(g\sigma f)}_2-
 \g {(g\sigma f)}_1{(g\alpha f)}_2
\nonumber\\
&=&-{(g\alpha f)}_{12}.
\end{eqnarray}
Here in the second line, antisymmetry of $\al _k$ ($k=1,2$)
(Eq.\ref{eq:antisym}) and the commutativity of $\sg _k$
(Eq.\ref{eq:sym}) have been used.

Proceeding similarly we have
\begin{theoremdemo}
  $\sg _{12}$ is symmetric.
\end{theoremdemo}
\begin{theoremdemo}
  $\alpha _{12}$ satisfies the Jacobi identity
\end{theoremdemo}
{\em Proof}:
\begin{eqnarray}
{((f\al g)\al h))}_{12}&=& \left[ \f {(f\al g)}_1{(f\sg g)}_2
+  \g {(f\sg g)}_1{(f\al g)}_2\right]\al h_{12}\nonumber\\
&=& \f\left[ \f{((f\al g)\al h)}_1{((f\sg g)\sg h)}_2+ \g
{((f\al g)\sg h)}_1{((f\sg g)\al h)}_2\right]\nonumber\\
\quad &+& \g\left[ \f{((f\sg g)\al h)}_1{((f\al g)\sg h)}_2
+ \g {((f\sg g)\sg h)}_1{((f\al g)\al h)}_2\right]\nonumber\\
&=&-\frac{1}{a_{12}}\left[
{(\Delta _\sg (g,h,f))}_1{((f\sg g)\sg h)}_2+
 {((f\sg g)\sg h)}_1{(\Delta _\sg (g,h,f))}_2\right]\nonumber\\
\quad &+&\frac{\sqrt{a_1a_2}}{a_{12}} {((f\al g)\sg h)}_1
\left\{\f {(f\al h)\sg g)}_2+\g {(f\sg (g\al h))}_2\right\}\nonumber\\
\quad &+&\frac{\sqrt{a_1a_2}}{a_{12}}\left\{\f {(f\al h)\sg g)}_1
+\g {(f\sg (g\al h))}_1\right\}{((f\al g)\sg h)}_2.
\end{eqnarray}
Here in the third equality, the CR (\ref{eq:asso}) has been used in
the first two terms and the derivation property (\ref{eq:der}), in the
last two terms.  Cyclically permuting $f$,$g$,$h$ in this relation
leads to two similar relations summing which results in separate
cancellation of all terms with the prefactor $-\frac{1}{a_{12}}$ and
those with the prefactor $\frac{\sqrt{a_1a_2}}{a_{12}}$, thus proving
the Jacobi identity.

We state two other lemmas the proofs of which are relegated to the
appendix because they are lengthy.
\begin{theoremdemo}
  $\alpha _{12}$ is a derivation wrt $\sg _{12}$, {\em i\ e\ .} the identity
  \mbox{(\ref{eq:der})} is satisfied in the TP space.
\end{theoremdemo}
\begin{theoremdemo}
  $\sg _{12}$ and $\alpha _{12}$ are related by the CR
  \mbox{(\ref{eq:asso})}.
\end{theoremdemo}
Now in view of the lemmas 1-5, we have \newtheorem{theorems}{Theorem}
\begin{theorems}
  The algebra ${\cal H}^{a_{12}}$ is a QHA.
\end{theorems}
This is our main result. This result can be interpreted in the
  following way. Suppose one starts with
  two physical systems each describable by its Hamiltonian and its own
  collection of observables satisfying the evolution given by its Lie
  bracket $\alpha$ and satisfying the properties expressed by the
  identities (\ref{eq:antisym}-\ref{eq:asso}). The two systems may
  require different PC's for their complete (algebraic) description.
  Yet their composite is describable by an `interaction' Hamiltonian
  along with its other observables following the evolution by a Lie
  bracket and also satisfying the same identities
  (\ref{eq:antisym}-\ref{eq:asso}) and with a PC which is in principle
  different from the PC's associated with the components.  Thus two
  quantum systems with different PC's can in principle interact in a
  scheme which provides for a consistent dynamics. We shall however
show that another natural algebraic requirement restricts this
possibility further.  Implications of this theorem is described in the
next section.

Before ending this section, we note the composition laws of $\sg$ and
$\al$ products for the special case for which the component HA's and
their TP are all characterized by the {\em same} quantum constant $a$:
\begin{eqnarray}
\label{spclcompsg}
{(f\sigma g)}_{12} &=& {(f\sigma g)}_1\otimes {(f\sigma g)}_2-a\,\,{(f\alpha
  g)}_1\otimes {(f\alpha g)}_2,\\
\label{spclcompal}
{(f\alpha g)}_{12} &=& {(f\alpha g)}_1\otimes
    {(f\sigma g)}_2+{(f\sigma g)}_1\otimes
    {(f\alpha g)}_2.
\end{eqnarray}
These laws were earlier derived in refs.\cite{grginpetersen,sahoo1}.
We also note two interesting facts: (a) whereas the composition law of
$\sg$ alone depends on $a$, that of $\al$ is {\em independent} of $a$
and (b) The composition law for the $\al^0$ product in a CHA is {\em
  identical} to that of the $\al^a$ product in a QHA. We now turn to
the treatment of mixed quantum and classical HA's.

\section{Mixed quantum-classical Hamilton algebra}
 A hybrid quantum-classical system is of considerable interest
  from the point of view of quantum measurement theory. In the
  orthodox Copenhagen philosophy of measurement, in order to measure
  an observable pertaining to a quantum system, one has to couple a
  classical system (the apparatus) with it for a certain duration of
  time during which the measurement takes place and subsequently the
  latter is decoupled from the former. A measurement is achieved if
  {\em unambiguous} information concerning the value of the measured
  variable is transfered (or ``stored'') into some suitable observable
  of the apparatus and thus one obtains this information (the
  ``pointer reading'') after the decoupling is over. This transfer
  process is technically referred to as the back-reaction of the
  quantum system. Theoretically one achieves it by using a coupling
  Hamiltonian in the TP space of observables of both the systems. It
  is but natural to look for a Lie product in the TP space which must
  be constructed out of the Lie products of the component algebras,
  one of which is the commutator bracket of operators (for the quantum
  system) and the other, the Poisson bracket of the phase-space
  functions (for the classical system).  Rather than banking upon the
  {\em correct} relation (\ref{newcompal}), one is tempted to be
  guided by the {\em inappropriate} relation
  (\ref{spclcompal}). In this relation, if
  one regards the component $1$ as being quantum and $2$ being
  classical, one would be naturally led to the mixed bracket
  (\ref{BT}) with appropriate algorithmic identifications. Note that
  from our point of view, this would be an {\em illegal} procedure not
  permitted in our analysis.  Misuse of (\ref{spclcompal}) is the
  reason why the bracket (\ref{BT}) and equivalently, the bracket
  (\ref{BT1}) does {\em not} satisfy the Jacobi identity--a fact which
  has been explicitly checked in reference\cite{carosalcedo}. It is
  the presence of the second term in the rhs of Eq.(\ref{spclcompal})
  which affects change in the classical variable. This is precisely
  the term responsible for the back-reaction of the system $1$
  (quantum) on the system $2$ (classical).
  
  A simple example to illustrate the back-reaction concept can be
  given for more clarification. Consider two quantum free particle
  systems labeled $k=1,2$ with masses $m_k$ and momenta operators
  $\hat{p}_k$.  Their Hamiltonians are given by
  $\hat{h}_k={\hat{p}_k}^2/2m_k$.  Let us be interested in measuring
  $\hat{p}_1$. A convenient coupling Hamiltonian for this purpose is
  $\hat{h}_{12}=g(t)\hat{p}_1\otimes \hat{x}_2$ where $\hat{x}_2$ is
  the position variable of the second particle and $g(t)$ is a
  coupling parameter such that it is everywhere zero, except between
  $t_0$ and $t_0+\Delta t$ (the duration of measurement), where it is
  constant (=$g_0$). Confirming to our notation, we have the full
  Hamiltonian
\begin{equation}
\hat{h}_{12}=\frac{\hat{p}_1^2}{2m_1}\otimes \hat{I}_2 + 
\hat{I}_1\otimes\frac{\hat{p}_2^2}{2m_2} + g(t)\hat{p}_1\otimes
\hat{x}_2.
\end{equation} 
Here $\hat{I}_k$ is the unit element of the QHA ${\cal H}^k$.
Following are the equations of motion, dictated by
Eq.(\ref{spclcompal}):
\begin{eqnarray}
\dot{\hat{p}}_1 &=& 0,\nonumber\\
\dot{\hat{x}}_1 &=& \frac{\hat{p}_1}{m_2}+g(t)\hat{x}_2,\nonumber\\
\dot{\hat{p}}_2 &=& -g(t)\hat{p}_1,\nonumber\\
\dot{\hat{x}}_2 &=& \frac{\hat{p}_2}{m_2} 
\label{eom}
\end{eqnarray}
Here $t$ is the time parameter and the time derivative is denoted by
an overdot.  We treat the system $1$ as one whose momentum $\hat{p}_1$
is to be measured and the system $2$ as the (quantum) apparatus.  The
first equation implies that $\hat{p}_1$ does not change as a result of
the measurement.  On solving for $\hat{p}_2$, we obtain
\begin{equation}
\hat{p}_2-\hat{p}_2^0=-g_0\hat{p}_1\Delta t.
\label{pmp0}
\end{equation}
This equation implies a correlation between $\hat{p}_2-\hat{p}_2^0$
and $\hat{p}_1$, such that if $\hat{p}_2-\hat{p}_2^0$ is observed, one
can calculate $\hat{p}_1$.  This change in momentum of the second
system arises due to the back-reaction of the first system. This
illustrates how the information transfer between the system and the
apparatus takes place in the von Neumann theory\cite{vonneumann} of
measurement.

We now consider strictly the Bohr point of view that the apparatus is
got to be a classical system. Suppose that in the above example the
system $1$ is quantum, but the system $2$ is classical. then it is no
longer correct to use Eq.(\ref{spclcompal}). Instead, one should use
relation (\ref{newcompal}).  Let $a_1=\hbar_1^2/4$, $a_2=0$ and let
$a_{12}=\hbar_{12}^2/4$. We then obtain immediately from
Eqs.(\ref{eq:sig12},\ref{eq:al12}) the results for the mixed products:
\begin{eqnarray}
\sg_{12} &=&(\sg _1\otimes\sg _2)\circ S,\\ 
\al _{12} &=& \frac{\hbar}{\hbar _{12}}\, (\al _1\otimes\sg _2)\circ S.
\end{eqnarray}
The constant $\hbar _{12}$ being arbitrary, we have to decide on its
value from some other consideration. The value $\hbar _{12}=0$ does
not lead to any meaningful structure whereas the choice $\hbar
_{12}=\hbar$ would mean that ${\cal H}^a\otimes{\cal H}^0$ ($a=\hbar
^2/4$) is a QHA. It would mean that a mixed quantum-classical system
is a {\em quantum} system governed by a Lie bracket $\al_{12}=(\al
_1^a\otimes\sg _2^0)\circ S$.  We therefore have
\begin{eqnarray}
\label{hybal}
(f_1\otimes f_2)\al_{12}(g_1\otimes g_2)={(f\al^ag)}_1\otimes {(f\sg^0g)}_2,
\\
\label{hybsg}
(f_1\otimes f_2)\sg_{12}(g_1\otimes g_2)={(f\sg^ag)}_1\otimes {(f\sg^0g)}_2.
\end{eqnarray}
In algorithmic form, these brackets are
\begin{eqnarray}
{[\{{Xx,Yy}\}]}^- &=& {[X,Y]}^-xy,\\
{[\{{Xx,Yy}\}]}^+ &=& {[X,Y]}^+xy.
\end{eqnarray}
Comparison Eqs. (\ref{hybal}) and (\ref{spclcompal}) reveals that the
second term in the rhs of (\ref{spclcompal}) is no longer present in
the rhs of (\ref{hybal}). Following through the example of measurement
given above it is clear that the absence of the product $\al_2$ in the
composition law (\ref{hybal}) leads to ``freezing'' of classical
dynamics. In other words there is no back-reaction effect of the
quantum system on the classical.  
This is a {\em no go} result.
Clearly, it does {\em not} mean that interactions between classical
and quantum systems vanish. Interactions exist and are given by
elements belonging to the tensor product algebra ${\cal
  H}^a\otimes{\cal H}^0$. These, however, do {\em not} affect any
change in the classical ``pointer'' variable!  This pinpoints the root
cause of the impediment to a satisfactory description of dynamical
evolution of a mixed quantum-classical system.  No wonder that the
standard Lie bracket\cite{anderson,bouchertraschen,prezhdokisil} for
mixed quantum-classical system, suggested along the line of
Eq.(\ref{spclcompal}), is {\em not} compatible with the algebraic
requirement. This explains the futility of such approaches.  We now
turn to another important result.

\section{Uniqueness of Planck's constant}
In the standard definition\cite{greub} of TP of two (associative)
algebras $\{A_k,\mu _k,\mathbb{F}\}$, ($k=1,2$) where $\mu _k$ is the
product law and $\mathbb{F}$ is the field, one has the composition law
$$
\mu _{12}=(\mu _1\otimes\mu _2)\circ S
$$
defined in the TP space $A_1\otimes A_2$. Let $e_k$ be the unit
element of $A_k$, {\em i.e.}, ${(f\mu e)}_k={(e\mu f)}_k=f_k$. Restriction
of the product $\mu _{12}$ to the component $A_1$, denoted $\mu
_{12}\mid A_1$ should result in the product $\mu _1\otimes e_2$. This
means, for two elements $f_1\otimes e_2$, $g_1\otimes e_2\in
A_1\otimes e_2$, we have
\begin{eqnarray}
(f_1\otimes e_2)\mu _{12}(g_1\otimes e_2)
&=&{(f\mu g)}_1\otimes {(e\mu e)}_2\nonumber\\
&=&{(f\mu g)}_1\otimes e_2.
\end{eqnarray}
Similarly, $\mu _{12}\mid A_2=e_1\otimes\mu _2$.

Extending the above restriction requirement to the HA, we note that
$\al $ being a Lie product, does not have an unit element and the Lie
product of any element with $e$ (the unit element of the $\sg $
product) vanishes, {\em i.e.}, $f\al e=0$ Algebraically, this means that the
HA is {\em central}, {\em i.e.}, if $f\al x=0$ holds for arbitrary $f\in
{\cal H}$, then $x=c\sg e$ where $c\in\mathbb{R}$.  We are now in a
position to prove \newtheorem{theoremdemo6}{Theorem 2}
\begin{theorems}
  There can be only one PC.
\end{theorems}
{\em Proof}: Consider the TP ${\cal A}^{a_{12}}$ of the associative
enveloping algebras of two QHA's ${\cal A}^{a_1}$ and ${\cal
  A}^{a_2}$.  Then we have
\begin{eqnarray}
\tau _{12}\mid A_1=\tau _1\otimes e_2,\nonumber\\
\tau _{12}\mid A_2=e_1\otimes\tau _2.
\label{eq:tau12}
\end{eqnarray}
Using the relations (\ref{eq:taucomp},\ref{eq:sig12},\ref{eq:al12})
and equating the symmetric and antisymmetric parts separately, we
obtain
\begin{eqnarray}
\al _{12}\mid A_1=\f\,(\al _1\otimes e_2),\nonumber\\
\al _{12}\mid A_2=\g\,(e_1\otimes\al _2).
\label{eq:alpf}
\end{eqnarray} 
wherein we have used the relation ${(e\sg e)}_k=e_k$. We now invoke
the standard restriction requirements:
\begin{eqnarray}
\al _{12}\mid A_1=(\al _1\otimes e_2),\nonumber\\
\al _{12}\mid A_2=(e_1\otimes\al _2).
\label{eq:rr}
\end{eqnarray}
Comparison of Eqs.(\ref{eq:alpf}) and (\ref{eq:rr}) leads to the
result
\begin{eqnarray}
&& a_{12}=a_1=a_2=a\quad\mbox{(say)}
\end{eqnarray}
and hence,
\begin{eqnarray}
&&\hbar _{12}=\hbar _1=\hbar _2=\hbar.
\end{eqnarray}
 The result that PC is unique, \marginpar{\em referee 1} apart
  from reaffirming the conventional assumption, also demonstrates
  clearly the very consistency of the HA approach.
\section{Conclusion}
To summarize, we have followed an algebraic approach to QM which is
free of the position-momentum generators satisfying the Heisenberg
commutation relation. It gives importance to the CR (\ref{eq:asso})
relating both the Lie and the Jordan products. In this way the concept of a
QHA subsumes symplectic ({\em i.e.}, position-momentum generated
algebra) as well as nonsymplectic ({\em i.e.}, algebras for internal degrees
of freedom such as spin) variables. We first show that two QHA's with
different PC's can lead to a composite QHA with yet another PC. We
then show that a QHA can form a composite with a CHA resulting in a
QHA and thus allowing for the only consistent description of a mixed
mechanics. However, back-reaction of the quantum system on the
classical is shown to be ruled out in this scheme. This result has
important bearings on quantum measurement issues as there is no 
way to describe a quantum system and a (classical) measuring apparatus
in a consistent way in the sense of Bohr. We have also proved, based
on the natural restriction requirement, that there is but one PC.

f In the light of our first result regarding freezing of dynamics
  in the classical-quantum system interaction, one may wonder how the
  model such as the one proposed recently by Belavkin\cite{belavkin}
  is able to deal with the measurement
  problem. The answer to this puzzle lies in the simple fact that in
  that model (as in the original von Neumann\cite{vonneumann}) the
  apparatus is also assumed to have a wave function (or a wave packet)
  thus endowing it with quantum property.  The model is based on
  Schr\"{o}dinger picture and also one considers only the Lie
  evolution (dynamics) mediated by stochastic interaction and
  dissipation. The Jordan product ($\sigma ^a$ in our notation) of
  observables does {\em not} enter at all in the treatment of
  ref.\cite{belavkin}. For example, given two observables $f$ and $g$,
  their ``observable'' product $f\sg ^ag$ evolves (say, under a
  Hamiltonian $h$) according to the product $h\al ^a(f\sg ^ag)$. So
  also the observables $f$ and $g$ evolve under the same $h$.
  Consistency of time evolution requires the derivation law
  (\ref{eq:der}) to hold. Requirements such as this need to be
  satisfied by the observables of the coupled systems also. It is not
  possible to make such consistency checks in the Schr\"{o}dinger
  picture adopted in ref.\cite{belavkin}.

l The concept of HA as introduced here pertains to Bose systems.
  There is a Fermionic counterpart. It has
  been introduced in ref\cite{sahoo}.  In this case one needs a {\em
    graded} algebra structure and the identities defining the algebra
  are graded versions of the identities
  (\ref{eq:antisym}-\ref{eq:asso}). Results derived in the present
  work can be extended to Fermi HA's. This extension along with some
  results concerning composition of Bose and Fermi Hamilton algebras
  will be dealt with in a separate work.

\acknowledgments The author wishes to thank the Director, Institute
of Physics, Bhubaneswar, for his kind hospitality. He would also like
to express his thanks to Dr Arun Kumar Pati for his comments on the
manuscript.

\appendix
\section{Proof of lemma 4}
We compute first the lhs of the identity:
\begin{eqnarray}
&&{(f\al (g\sg h))}_{12}=f_{12}\al _{12}\left\{ {(g\sg h)}_1
{(g\sg h)}_2-\sqrt{a_1a_2}{(g\al h)}_1{(g\al h)}_2\right\}\nonumber\\
&&=\f {(f\al (g\sg h))}_1{(f\sg (g\sg h))}_2
 +\g {(f\sg (g\sg h))}_1{(f\al (g\sg h))}_2\nonumber\\
&&\quad -a_1\g {(f\al (g\al h))}_1{(f\sg (g\al h))}_2+
   a_2\f {(f\sg (g\al h))}_1{(f\al (g\al h))}_2\nonumber\\
&&=\f {((f\al g)\sg h)}_1{(f\sg (g\sg h))}_2
+\f {(g\sg (f\al h))}_1{(f\sg (g\sg h))}_2\nonumber\\
&&\quad +\g  {(f\sg (g\sg h))}_1{((f\al g)\sg h)}_2
+\g {(f\sg (g\sg h))}_1{(g\sg (f\al h))}_2\nonumber\\
&&\quad -a_1\g {((f\al g)\al h)}_1{(f\sg (g\al h))}_2
+a_1\g {(g\al (f\al h))}_1{(f\sg (g\al h))}_2\nonumber\\
&&\quad -a_2\f  {(f\sg (g\al h))}_1{((f\al g)\al h)}_2
+a_2\f {(f\sg (g\al h))}_1{(g\al (f\al h))}_2\nonumber\\
\label{eq:derlhs}
\end{eqnarray}  
The two terms on the rhs of the identity are:
\begin{eqnarray}
&&{(g\sg (f\al h))}_{12}=g_{12}\sg _{12}\left[\f {(f\al h)}_1{(f\sg h)}_2
+\g {(f\sg h)}_1{(f\al h)}_2\right]\nonumber\\
&&=\f {(g\sg (f\al h))}_1{(g\sg (f\sg h))}_2
+  \g {(g\sg (f\sg h))}_1{(g\sg (f\al h))}_2\nonumber\\
&&\quad - a_1\g {(g\al (f\al h))}_1\left[ 
{((g\al f)\sg h)}_2+{(f\sg (g\al h))}_2\right]\nonumber\\
&&\quad -a_2\f \left[ {((g\al f)\sg h)}_1+{(f\sg (g\al h))}_1\right]
{(g\al (f\al h))}_2
\label{eq:derrhs1}
\end{eqnarray}
and
\begin{eqnarray}
&&{((f\al g)\sg h)}_{12}=\left[ \f {(f\al g)}_1{(f\sg g)}_2+
\g {(f\sg g)}_1{(f\al g)}_2\right] \sg _{12} h_{12}\nonumber\\
&&=\f {((f\al g)\sg h)}_1{((f\sg g)\sg h)}_2+
\g {((f\sg g)\sg h)}_1{((f\al g)\sg h)}_2\nonumber\\
&&\quad -a_1\g {((f\al g)\al h)}_1\left[ {((f\al h)\sg g)}_2
+{(f\sg (g\al h))}_2\right]\nonumber\\
&&\quad -a_2\f \left[ {((f\al h)\sg g)}_1+{(f\sg (g\al h))}_1\right]
{((f\al g)\al h)}_2.
\label{eq:derrhs2}
\end{eqnarray}
We now simplify the combination
\begin{eqnarray}
&&{(f\al (g\sg h))}_{12}-{(g\sg (f\al h))}_{12}-{(f\al g)\sg h)}_{12}
\nonumber\\
&&=-\f {((f\al g)\sg h)}_1{(\Delta _\sg (f,g,h))}_2
   -\g {(\Delta _\sg (f,g,h))}_1 {((f\al g)\sg h)}_2\nonumber\\
&&\quad + \f {(g\sg (f\al h))}_1{(\Delta _\sg (g,h,f))}_2
  + \g {(\Delta _\sg (g,h,f))}_1 {(g\sg (f\al h))}_2\nonumber\\
&&\quad +a_1\g {((f\al g)\al h)}_1{((f\al h)\sg g)}_2
  +a_2\f {((f\al h)\sg g)}_1{((f\al g)\al h)}_2\nonumber\\
&&\quad +a_1\g {(g\al (f\al h))}_1{((g\al f)\sg h)}_2
  +a_2\f {((g\al f)\sg h)}_1{(g\al (f\al h))}_2.\nonumber\\
\end{eqnarray}
Making use of the CR (\ref{eq:asso}) in the associators occurring in
the above relation, we immediately see that the rhs vanishes thereby
establishing the derivation identity.

\section{Proof of lemma 5}
We first simplify the individual terms of the relation.
\begin{eqnarray}
&&{((f\sg g)\sg h)}_{12}
\nonumber\\
&&=\left[ {(f\sg g)}_1{(f\sg g)}_2-\s12 \,{(f\al g)}_1{(f\al
    g)}_2\right]\sg h_{12}
\nonumber\\
&&=\ssl fgh1\,\ssl fgh2-\s12\,\sa fgh1\,\sa fgh2
\nonumber\\
&&\quad -\s12\,\as fgh1\,\as fgh2+a_1a_2\,\aa fgh1\,\aa fgh2
\nonumber\\
&&=\ssl fgh1\,\ssl fgh2
\nonumber\\
&&\quad -\s12\,\left\{\as fhg1+\sar fgh1\right\}\left\{\as fhg1+\sar
  fgh2\right\}
\nonumber\\
&&\quad -\s12\,\as fgh1\,\as fgh2+{(\Delta _\sg (g,h,f))}_1\,{(\Delta _\sg
  (g,h,f))}_2
\nonumber\\
&&=\ssl fgh1\,\ssl fgh2-\s12\,\as fhg1\,\as fhg2
\nonumber\\
&&\quad -\s12\,\as fhg1\,\sar fgh2-\s12\,\sar fgh1\,\as fhg2
\nonumber\\
&&\quad -\s12\,\sar fgh1\,\sar fgh2-\s12\,\as fgh1\,\as fgh2
\nonumber\\
&&\quad +{(\Delta _\sg (g,h,f))}_1\,{(\Delta _\sg
  (g,h,f))}_2
\nonumber\\
&&=\ssl fgh1\,\ssl fgh2+\ssl ghf1\,\ssl ghf2
\nonumber\\
&&\quad -\ssl ghf1\,\ssr ghf2-\ssr ghf1\,\ssl ghf2
\nonumber\\
&&\quad +\ssr ghf1\,\ssr ghf2
\nonumber\\
&&\quad -\s12\,\as fhg1\,\as fhg2-\s12\,\as fhg1\,\sar fgh2
\nonumber\\
&&\quad -\s12\,\sar fgh1\,\as fhg2-\s12\,\sar fgh1\,\sar fgh2
\nonumber\\
&&\quad -\s12\,\as fgh1\,\as fgh2.
\end{eqnarray}
Here in the second equality, the derivation relation (\ref{eq:der})
and the CR (\ref{eq:asso}) have been used and in the fourth equality
the associator expressions have been explicitly substituted using
(\ref{eq:asso}).  Similar simplification steps lead to the result
\begin{eqnarray}
&&\ssr fgh{12}
\nonumber\\
&&=\ssr fgh1\,\ssr fgh2+\ssl hfg1\,\ssl hfg2
\nonumber\\
&&\quad -\ssl hfg1\,\ssr hfg2-\ssr hfg1\,\ssl hfg2
\nonumber\\
&&\quad +\ssr hfg1\,\ssr hfg2
\nonumber\\
&&\quad -\s12\,\as fgh1\,\as fgh2-\s12\,\as fgh1\,\sar gfh2
\nonumber\\
&&\quad -\s12\,\sar gfh1\,\as fgh2-\s12\,\sar gfh1\,\sar gfh2
\nonumber\\
&&\quad -\s12\,\sar fgh1\,\sar fgh2.\nonumber\\
\end{eqnarray}
The remaining term of the CR is
\begin{eqnarray}
&&a_{12}\aa fhg{12}
\nonumber\\
&&=a_{12}\left[\f\,{(f\al h)}_1{(f\sg h)}_2+\f\,{(f\al h)}_1{(f\sg
    h)}_2\right]\al _{12}(g_{12})
\nonumber\\
&&=a_{12}\,\f\left[\f\,\aa fhg1\,\ssl fhg2+\g\,\as fhg1\,\sa
  fhg2\right]
\nonumber\\
&&\quad +a_{12}\,\g\,\left[\f\,\sa fhg1\,\as fhg2+\g\,\ssl fhg1\,\aa
  fhg2\right]
\nonumber\\
&&=a_1\,\aa fhg1\,\ssl fhg2+\s12\,\as fhg1\,\left[\as fgh2+\sar
  fhg2\right]
\nonumber\\
&&\quad +\s12\,\left[\,\as fgh1+\sar fhg1\right]\,\as fhg2
\nonumber\\
&&\quad +a_2\,\ssl fhg1\,\aa fhg2
\nonumber\\
&&={(\Delta _\sg (f,g,h))}_1\,\ssl fhg2+\s12\,\as fhg1\,\as fgh2
\nonumber\\
&&\quad +\s12\,\as fhg1\,\sar fhg2+\s12\,\aa fgh1\,\as fhg2
\nonumber\\
&&\quad +\s12\,\sar fhg1\,\as fhg2+\ssl fhg1\,{(\Delta _\sg (f,g,h))}_2
\nonumber\\
&&=\ssl fgh1\,\ssl fhg2-\ssr fgh1\,\ssl fhg2
\nonumber\\
&&\quad +\ssl fhg1\,\ssl fgh2-\ssl fhg1\,\ssr fgh2
\nonumber\\
&&\quad +\s12\,\as fhg1\,\as fgh2+\s12\,\as fhg1\,\sar fhg2
\nonumber\\
&&\quad +\s12\,\as fgh1\,\as fhg2+\s12\,\sar fhg1\,\as fhg2.
\nonumber\\
\end{eqnarray}
Combining these three terms leads to the final result
\begin{eqnarray*}
&&\ssl fgh{12}-\ssr fgh{12}-a_{12}\aa fhg{12}=0,
\end{eqnarray*}
{\em i.e.},\quad\quad\quad\quad\quad\quad\quad $ {\Delta _\sg
  (f,g,h)}_{12}=a_{12}\,\aa fhg{12}.$

\end{document}